\documentclass[twocolumn,amsmath,floatfix,amssymb,aps,prl,showpacs,superscriptaddress]{revtex4-1}
\usepackage{graphicx,natbib}
\usepackage{amsmath}
\usepackage[colorlinks,bookmarks=false,citecolor=blue,urlcolor=blue]{hyperref}
\usepackage{color}
\usepackage{ulem}

\begin{document}

\title{Quantum phase transitions and decoupling of magnetic sublattices in the quasi-two-dimensional Ising magnet Co$_{3}$V$_{2}$O$_{8}$ in a transverse magnetic field}
\author{K. Fritsch}
\affiliation{Department of Physics and Astronomy, McMaster University, Hamilton, Ontario, L8S 4M1, Canada}
\affiliation{Helmholtz Zentrum Berlin f\"{u}r Materialien und Energie GmbH, D-14109 Berlin, Germany}

\author{G. Ehlers}
\affiliation{Quantum Condensed Matter Division, Oak Ridge National Laboratory, Oak Ridge, Tennessee 37831-6475, USA}

\author{K. C. Rule}
\altaffiliation[current address: ]{The Bragg Institute, ANSTO, Kirrawee DC NSW 2234, Australia}
\affiliation{Helmholtz Zentrum Berlin f\"{u}r Materialien und Energie GmbH, D-14109 Berlin, Germany}

\author{K. Habicht}
\affiliation{Helmholtz Zentrum Berlin f\"{u}r Materialien und Energie GmbH, D-14109 Berlin, Germany}

\author{M. Ramazanoglu}
\altaffiliation[current address: ]{Physics Eng. Dept., Science and Letters Faculty, Istanbul Technical University, Maslak, TR-34469, Istanbul, Turkey}
\affiliation{Department of Physics and Astronomy, McMaster University, Hamilton, Ontario, L8S 4M1, Canada}

\author{H. A. Dabkowska}
%\affiliation{Department of Physics and Astronomy, McMaster University, Hamilton, Ontario, L8S 4M1, Canada}
\affiliation{Brockhouse Institute for Materials Research, Hamilton, Ontario, L8S 4M1, Canada}

\author{B. D. Gaulin}
\affiliation{Department of Physics and Astronomy, McMaster University, Hamilton, Ontario, L8S 4M1, Canada}
\affiliation{Brockhouse Institute for Materials Research, Hamilton, Ontario, L8S 4M1, Canada}
\affiliation{Canadian Institute for Advanced Research, 180 Dundas St.\ W.,
Toronto, Ontario, M5G 1Z8, Canada}

%\date{\today}

\begin{abstract}

The application of a magnetic field transverse to the easy axis, Ising direction in the quasi-two dimensional Kagome staircase 
magnet, Co$_3$V$_2$O$_8$, induces three quantum phase transitions at low temperatures, ultimately producing a novel high field polarized state, with two distinct sublattices. New time-of-flight neutron scattering techniques, accompanied by large angular access, high magnetic field infrastructure allow the mapping of a sequence of ferromagnetic and incommensurate phases and their accompanying spin excitations. At least one of the transitions to incommensurate phases at $\mu_0H_{c1}\sim6.25$ T and $\mu_0H_{c2}\sim7$ T is discontinuous, while the final quantum critical point at $\mu_0H_{c3}\sim 13$ T is continuous.  
 
\end{abstract}
\pacs{75.25.-j,75.30.Kz,75.30.Ds,75.40.Gb}

\maketitle

%\section{Introduction}
Quantum phase transitions (QPTs) have attracted great interest due to their relevance to unconventional magnetism and superconductivity in heavy fermion systems\cite{Si2010,Sachdev2000} and high temperature superconductors\cite{Sachdev2000}. QPTs are driven by quantum mechanical, rather than thermal fluctuations, and these can be tuned by external parameters such as pressure \cite{RueggTlCuCl3}, magnetic field\cite{Sachdev2000,Gegenwart2002}, or chemical doping\cite{Coleman2000}. The Ising model in a transverse magnetic field generates one of the canonical examples of a system with a QPT. The generic Hamiltonian for such a transverse field Ising model can be written as:
\begin{equation}
\mathcal{H}=-\sum_{ij}J_{ij}S_{i}^{z}S_{j}^{z}-g\mu_{\rm B}H\sum_iS_i^x.
\end{equation}
Here, the $J_{ij}$ exchange term couples the $z$-components of the spin operators $S$ at sites $i$ and $j$, while the magnetic field $\mu_{\rm B}H$ acts on the \textit{transverse}, $x$-component of the spins only. The transverse field mixes the spin states and ultimately leads to the destruction of long range order at zero temperature and a critical field $H_c$. Two experimental realizations of this model have been well studied: the three dimensional dipolar-coupled uniaxial ferromagnet LiHoF$_4$\cite{Ronnow2005,Bitko1996}, and the quasi-one-dimensional exchange-coupled Ising ferromagnet CoNb$_2$O$_6$\cite{Coldea2010,Kinross2014,Cabrera2014}.

In this paper, we report on transverse field induced QPTs in the \textit{quasi-two dimensional} (2D) Ising-like system Co$_3$V$_2$O$_8$ (CVO). The transverse magnetic field competes directly against the tendency of intersite magnetic interactions to order, however few Ising magnets exist such that an experimentally accessible magnetic field in a neutron scattering experiment, capable of determining structure and dynamics, can overcome their relevant exchange interactions. The development of a 16 T magnet cryostat \cite{FatSam}, with large horizontal plane access, and modern time-of-flight neutron scattering techniques have allowed a comprehensive 4-dimensional neutron scattering study on CVO. This study was performed over a wide range of fields covering three QPTs, and yields a novel two-sublattice field-polarized state at transverse fields above H$_{c3}$.

\begin{figure}
\includegraphics[width=7cm]{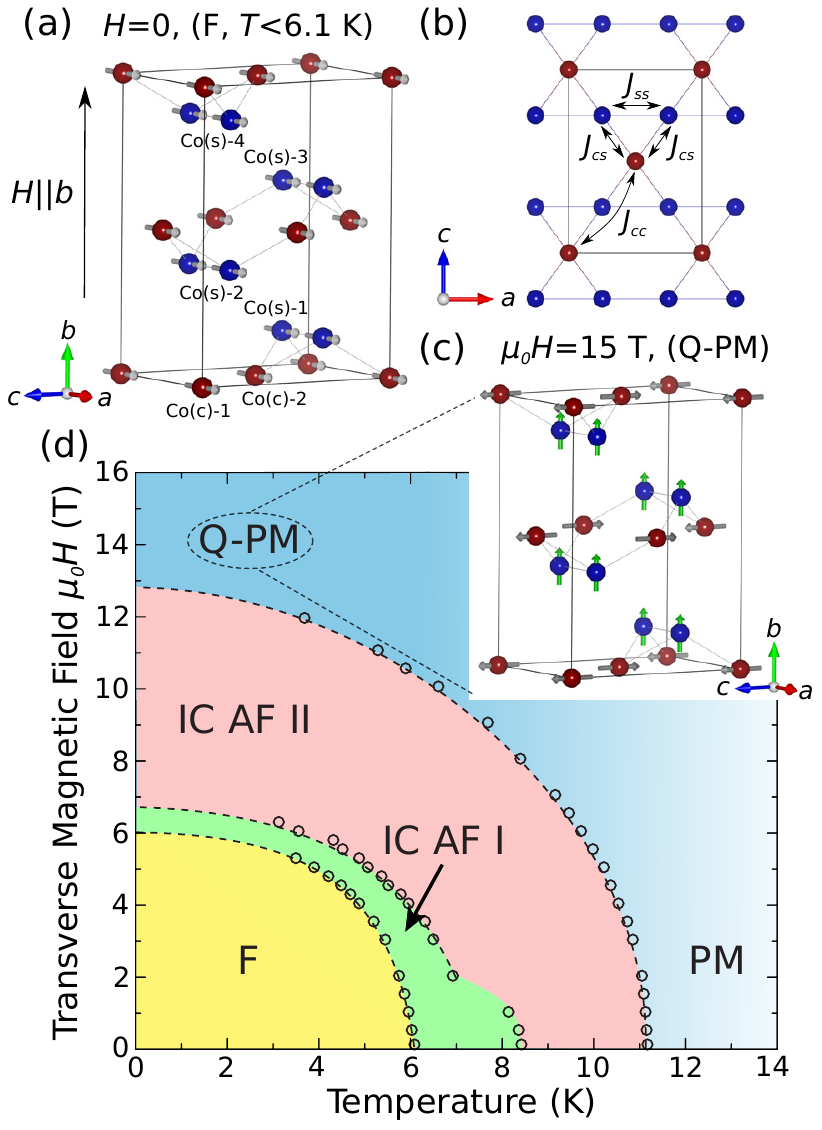}
\caption{(Color online) (a) View of the kagome staircase in CVO considering only the Co$^{2+}$ ions. The magnetic field is applied along the hard, $b$-direction, transverse to the magnetic easy axis ($a$). The ferromagnetic zero-field spin configuration is illustrated with grey arrows pointing along $a$. (b) View of the buckled kagome plane projected on the \textit{a}-\textit{c} plane with cross-tie (red) and spine sites (blue), and showing the relevant magnetic exchange interactions discussed in the text. (c) The high-field spin configuration in the Q-PM phase with the two decoupled spin sublattices as discussed in the text. (d) Magnetic phase diagram for $H\parallel b$ determined from magnetization measurements.  Above 13 T, the system is in a quantum paramagnetic (Q-PM) phase that connects continuously to the paramagnetic (PM) phase above 11.2 K in $H=0$.}
\label{fig:fig1}
\end{figure}

CVO is a member of the kagome staircase family of materials with formula M$_3$V$_2$O$_8$ (M=Co, Ni, Cu, Mn),\cite{Rogado02,RogadoCu3V2O8,Morosan} which are characterized by a stacked arrangement of buckled kagome layers, an anisotropic version of the 2D kagome lattice. In CVO, the magnetism arises from Co$^{2+}$ ions that occupy two crystallographically inequivalent sites within these layers, which give rise to a complex crystal field scheme that has yet to be analysed in detail \cite{WilsonPhD}. A representation of the magnetic structure is shown in Fig. \ref{fig:fig1}(a). Within the buckled kagome layers in the $a$-$c$ plane (Fig. \ref{fig:fig1}(b)), the Co$^{2+}$ magnetic moments on the spine sites (s) form chains along the $a$-axis, and these chains are linked by the cross-tie sites (c). The layers are well separated along the $b$-direction by nonmagnetic vanadium (V$^{5+}$) oxide layers, giving rise to the quasi-2D nature of this system. 

The low temperature phase diagram of CVO has been studied by neutron diffraction in both zero \cite{Wilson07,Chen06,Mehmet09,Qureshi_form_factor09} and finite applied magnetic fields \cite{Petrenko10,Helton2012}, by optical spectroscopy \cite{Rai,Vergara_magnetoelastic_10}, heat capacity and magnetization \cite{Rogado02,Szymczak06} as well as $\mu$SR \cite{muons07} and NMR \cite{NMRCVO_NVO,VNMRCVO} measurements. In zero field, CVO displays four different incommensurate and commensurate antiferromagnetic phases below 11.2 K that ultimately terminate in a ferromagnetic ground state below $T\sim$6.2 K. All five of the magnetic states display a preferred direction of the spins parallel to the \textit{a}-axis, the easy axis of this system. The phase diagram is highly anisotropic, and the susceptibility in the ferromagnetic state along the easy $a$-axis is almost two orders of magnitude larger than that along the hard $b$-axis \cite{Wilson07,dopedCVO,Szymczak06}. The susceptibility along $c$ is intermediate between that along $a$ and $b$, and we therefore expect CVO to approximate a quasi-2D Ising system. Within the ferromagnetic phase, the ordered moments on the cross-tie sites are $1.8\mu_{\rm B}$, almost a factor of two smaller than the $3\mu_{\rm B}$ moments on the spine sites \cite{Wilson07}. The spin Hamiltonian is approximately known \cite{Mehmet09} from earlier zero field neutron measurements which treated CVO as a 2D system and employed linear spin wave theory. The resulting exchange parameters between magnetic moments on the cross-tie and spine sites were found to be ferromagnetic with $J_{\mathrm{cs}}\sim1.25$ meV, while the exchange between adjacent spine sites ($J_{\mathrm{ss}}$ in Fig. \ref{fig:fig1}(b)) was found to be negligible. Uniaxial anisotropy terms on the order of $\sim$1-2 meV were also found, in agreement with the easy axis anisotropy. The influence of small magnetic fields ($\mu_0 H<2.5$ T) on the ground state of CVO has also been investigated for fields applied along the $a$-axis \cite{Helton2012} and along the $a$- and $c$-axes \cite{Petrenko10}. 

%\section{Experimental Details}
A high-quality single crystal of CVO grown using a floating zone image furnace \cite{HannaOFZ} was cut to a rectangular shape (18 x 5 x 5 mm$^3$) of mass $\sim$ 2.1 g and carefully aligned to less than 0.5$^{\circ}$ in the horizontal $(H,0,L)$ scattering plane for the neutron scattering measurements. Magnetization measurements were performed on a smaller $\sim$ 50 mg crystal  ($\sim$2.5 x 2 x 2 mm$^3$) cut from the same crystal boule. These measurements were performed at Helmholtz Zentrum Berlin, using a 14 T VSM option of a Quantum Design PPMS. Neutron scattering data were obtained using the cold time-of-flight spectrometer CNCS \cite{Ehlers2011} at the Spallation Neutron Source, Oak Ridge National Laboratory, employing $E_i=$12 meV neutrons with an elastic energy resolution of 0.45 meV. The magnetic field was supplied by the 16 T vertical field magnet cryostat ``FatSam" with a base temperature of 1.6 K.

Magnetization measurements were carried out with $H\parallel b$ over a temperature range of $\sim$3-15 K both as a function of temperature at fixed field and as a function of field at fixed temperature. Critical fields and temperatures were extracted using anomalies in the differential susceptibilities ${\rm d} M/{\rm d}T|_{\rm H_{fix}}$ and ${\rm d} M/{\rm d}H|_{\rm T_{fix}}$. The resulting phase diagram is shown in Fig. \ref{fig:fig1}(d). In agreement with previous reports, we find zero-field transitions at 11.2 K [paramagnetic (PM) to incommensurate antiferromagnetic (IC AF II)], at 8.3 K [IC AF II to IC AF I], and at 6.2 K [IC AF I to ferromagnetic (F)]. We largely reproduce the previously reported low field region of the phase diagram\cite{Wilson07,Yasui2007}. As a function of magnetic field, we observe three QPTs at 1.6 K. By extrapolating the higher temperature measurements to $T\sim$ 0, we find QPTs from the F to the IC AF phase I ($\sim 6.2$ T), the IC AF I to IC AFM II ($\sim 7$ T) and from IC AF II to a field polarized ``quantum paramagnetic" (Q-PM) phase at $\sim 13$ T. The term "quantum paramagnetic" is used here to distinguish this field polarized paramagnetic phase from the thermal paramagnetic phase at H=0 at $T\geq11.2$ K to which it is smoothly connected.

\begin{figure}
\includegraphics[width=8cm]{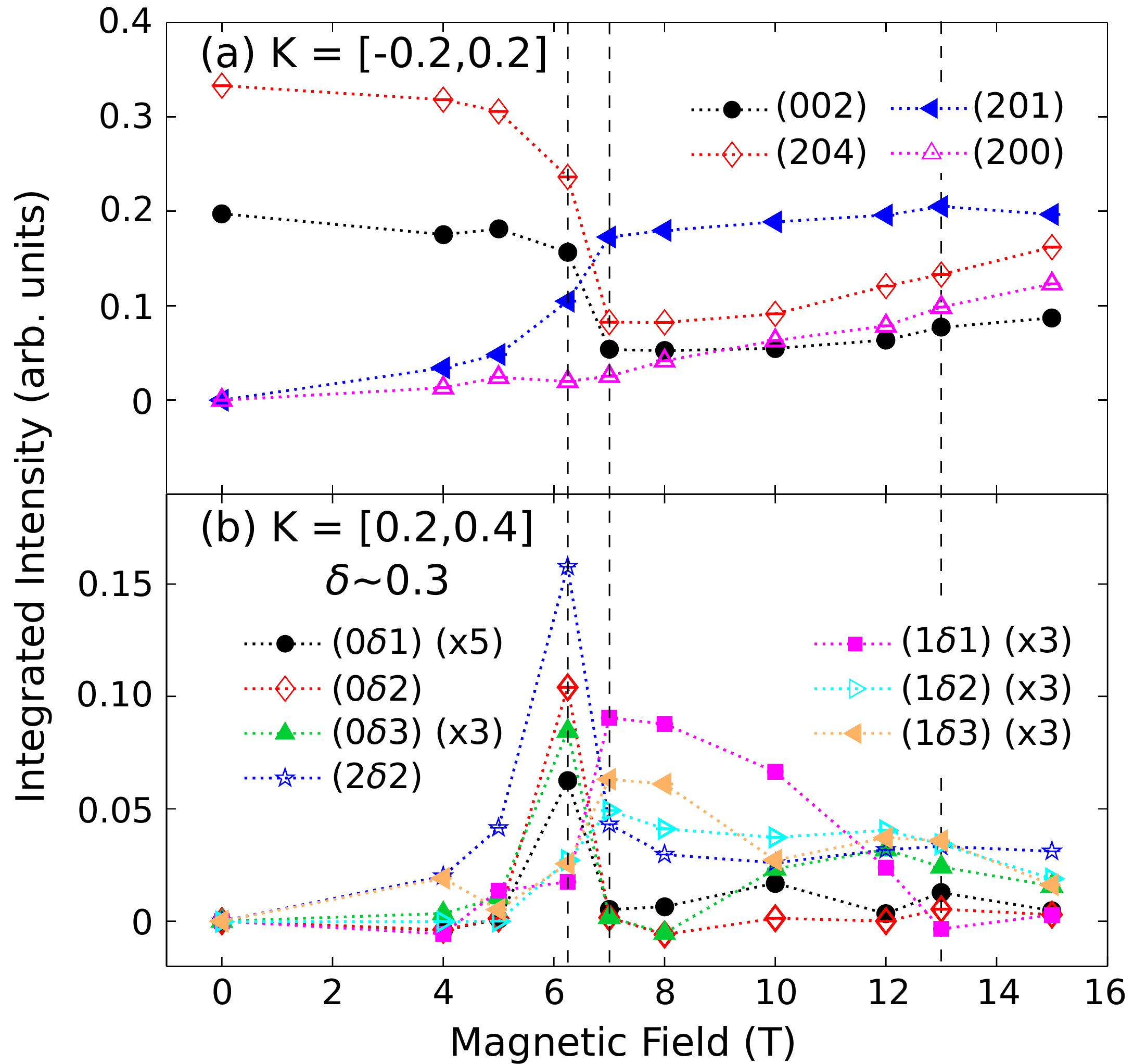}
\caption{Integrated areas of the elastic peaks shown in Fig. S1 for peak positions that appear (a) in the horizontal plane for $K = [-0.2, 0.2]$ r. l. u. and (b) above the horizontal plane for $K = [0.2, 0.4]$ r. l. u., giving an approximate incommensuration of $\delta\sim0.3$. QPTs at $\mu_0H_{\rm c1}\sim$ 6.25 T,  $\mu_0H_{\rm c2}\sim$ 7 T and $\mu_0H_{\rm c3}\sim$ 13 T (dashed lines) are clearly observable.}
\label{fig:intelasticpeaks} 
\end{figure}

Time-of-flight neutron scattering techniques allowed us to map out both the elastic and inelastic scattering function $S\left({\bf Q}, E\right)$ over a wide range of reciprocal space.
The transverse field dependence of the magnetic Bragg scattering ($-0.4<E<0.4$ meV) at $T=1.6$ K is shown in Fig. \ref{fig:intelasticpeaks}(a) and (b) for Bragg positions of type (H,0,L) and (H,$\delta\sim0.3$,L), respectively. These integrated intensities have had their 30 K analogue subtracted off to isolate the magnetic Bragg intensity. Reciprocal space maps of the elastic data sets are shown in the Supplemental Material (Fig. S1).  Zero-field magnetic scattering is strong at $(0,0,2)$ and $(2,0,4)$ positions indicating a F structure in which the moment size on the cross-tie sites is about half of the $\sim 3 \mu_{\rm B}$ moment at the spine sites. All moments are aligned along the $a$-axis. This is consistent with previous reports of the zero-field magnetic structure\cite{Chen06,Wilson07}. At $\mu_0H\parallel b =$ 6.25 T and 7 T, the incommensurate magnetic structure is complex and only partially understood on the basis of the data collected. It is consistent with a) a doubling of the unit cell along the $c$-direction and b) an incommensuration of $\delta\sim0.3$ along $b$ (see Fig. \ref{fig:intelasticpeaks}(b)). Including the strong $(2,0,1)$ peak and an absence of scattering at $(0,0,1)$, one deduces that a subset of the moments aligns antiferromagnetically along the $c$-axis. For transverse fields beyond 7 T, scattering at incommensurate wave vectors, in particular $(1,\delta,1)$, falls off continuously as the final QPT is approached at $\mu_0H_{\rm c3}\sim 13$ T.

Surprisingly, the elastic scattering at our highest transverse field, $\mu_0H\parallel b =$ 15 T, is not consistent with a simple polarized magnetic structure. Magnetic scattering at $(2,0,1)$ and $(2,0,0)$ along with an absence of scattering at $(0,0,1)$ imply an AF arrangement of the cross-tie moments which point along the $c$-axis. The spine-site moments, in contrast, are polarized along the $b$-axis in the transverse field direction. Within this two sublattice structure the scattering is consistent with the cross-tie sites having their full $\sim 3 \mu_{\rm B}$ moment, and the spine sites displaying only $\sim 1\pm0.2 \mu_{\rm B}$, less than half of their full, zero-field moment. The two spin sites (cross-tie and spine) then form two decoupled sublattices, both of which display moments normal to the easy-magnetization axis: along $c$ for the AF-correlated cross-tie sites, and along the hard, $b$-axis for the field-polarized spine sites, schematically shown in Fig. \ref{fig:fig1}(c). Such an AF cross-tie sublattice, flopped along $c$, requires AF exchange between neighbouring cross-tie moments, schematically indicated as $J_{\mathrm{cc}}$ in Fig. \ref{fig:fig1} (b). This competes with the dominant ferromagnetic $J_{\mathrm{cs}}$ interactions and is presumably what is responsible for the incommensurate phases at intermediate temperatures and fields. The decoupling between spine and cross-tie sublattices allows the ordered cross-tie sites to assume their full ordered moment within an AF structure.  This novel high field phase can then be continuously deformed into a fully polarized state by progressively canting the component of cross-tie moment along $b$, until both the cross-tie moments and the spine moments are fully polarized. No further QPT is required to achieve full polarization, and the $\mu_0H\parallel b =$ 15 T phase connects continuously to the paramagnetic phase in zero field.  

\begin{figure*}
\includegraphics[width=16cm]{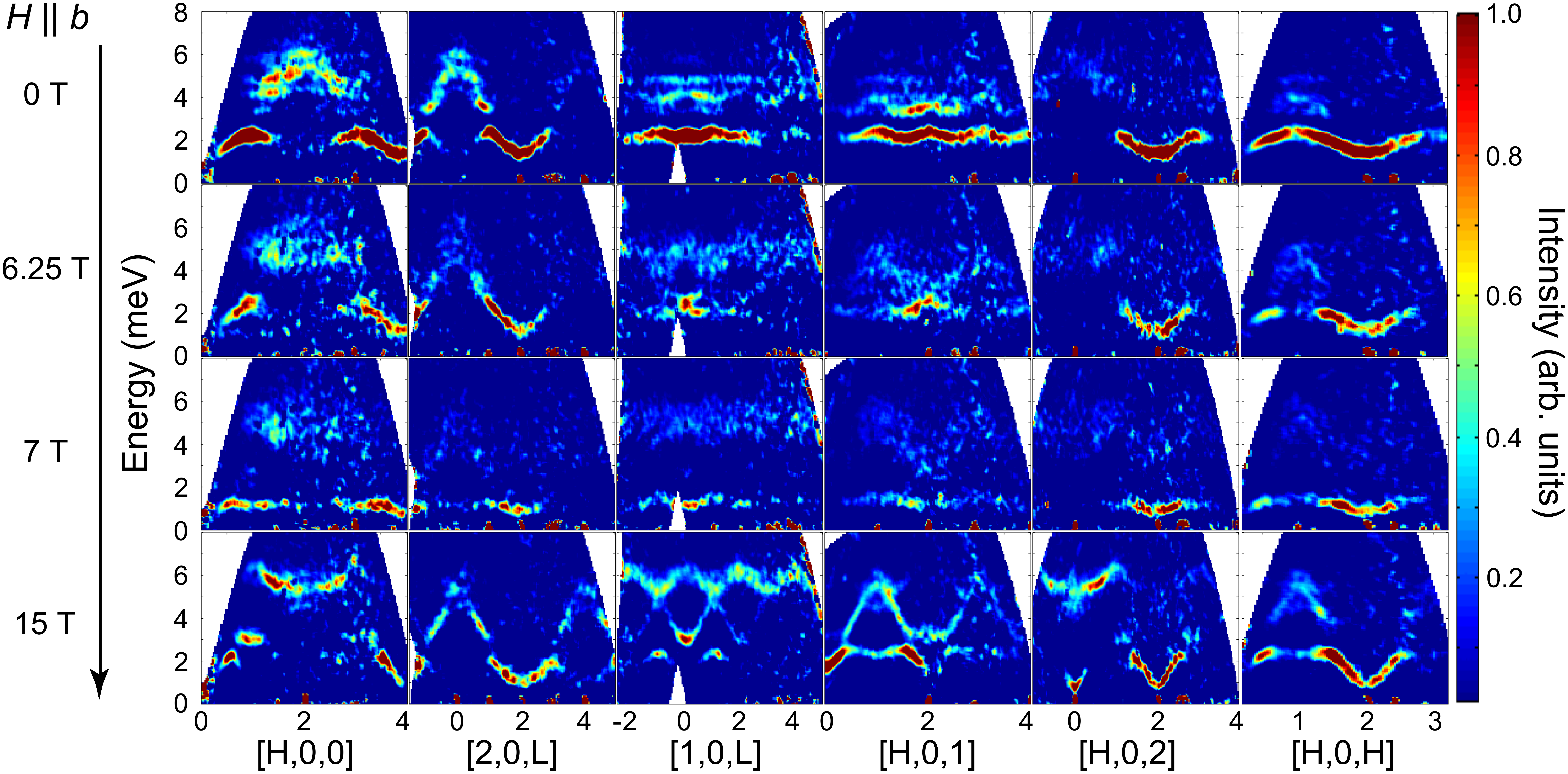}
\caption{Representative energy vs wave vector slices for various directions in reciprocal space, at $T=1.6$ K. The integration range for each of these sets is $\pm0.4$ r. l. u. in $K$ and $\pm0.25$ r. l. u. in either $H$ or $L$ direction. A $T=30$ K background is subtracted from the data. Measurements are shown in each of the $T=1.6$ K  phases at $\mu_0H\parallel b$ = 0, 6.25, 7 and 15 T.}
\label{fig:inelasticoverview}
\end{figure*}

Inelastic scattering maps $S({\bf Q}, E)$ are shown in Fig. \ref{fig:inelasticoverview} for several directions in reciprocal space, as a function of transverse magnetic field. All of these data have had the 30 K, zero field, background subtracted from them.
In zero field (top row of Fig. \ref{fig:inelasticoverview}), the spectrum is described by two relatively broad bands of spin wave scattering. The low energy band is centered near $\sim$ 2 meV with a $\sim1.4$ meV gap at $\bf{Q}$s such as (4,0,0) and (2,0,2).  The higher energy band is centered near 5 meV and is separated by $\sim$ 1 meV from the lower energy band at ${\bf Q}$s such as (3,0,0) and (2,0,1). Higher energy resolution measurements shown in the supplemental material (see Figs. S2 and S3) reveal the lower energy band to be comprised of two resolution limited spin wave modes, while the upper band is composed of four resolution limited modes, for a total of six spin wave modes, consistent with three magnetic sites per unit cell.

\begin{figure}[h]
\includegraphics[width=8cm]{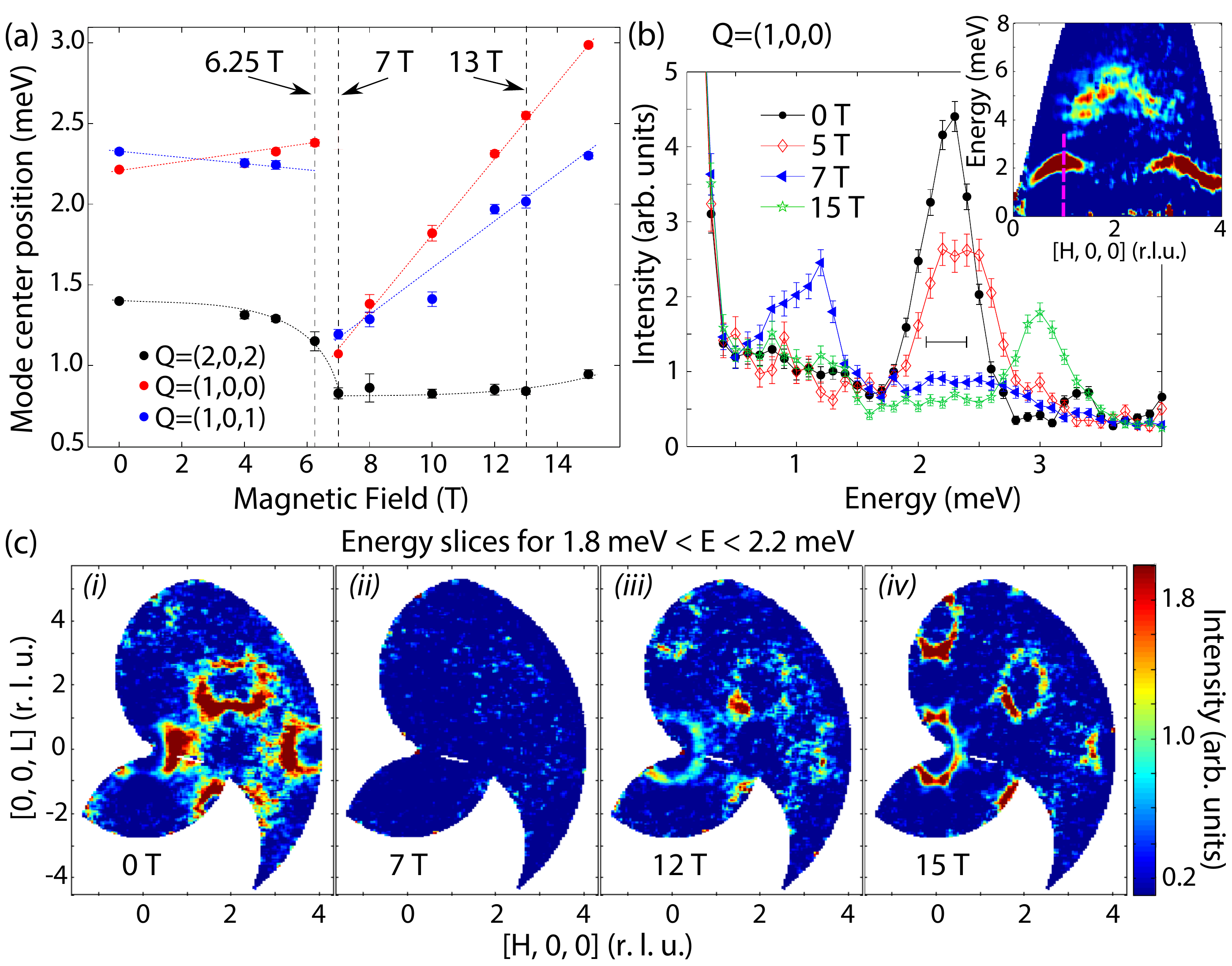}
\caption{(Color online) (a) Magnetic field dependence at $T=1.6$ K of the lowest-lying spin wave modes $E<$3.5 meV and the evolution of their energy gaps shown for three wavevectors, ${\bf Q}=(2,0,2)$, $(1,0,0)$, $(1,0,1)$. QPTs at $\mu_0H\sim$ 6.2 T, 7 T, and  13 T are indicated by dashed vertical lines. (b) Representative ${\bf Q}=(1,0,0)$ cut vs energy extracted from 2D data set (inset) with energy resolution indicated as horizontal bar. (c) Constant energy slices for $1.8<E<2.2$ meV in the $(H, 0, L)$ plane for $\mu_0H\parallel b$ = 0, 7, 12 and 15 T, cutting through the lowest lying spin wave dispersion surfaces.}
\label{fig:gapdependence}
\end{figure}

The spin excitation spectrum evolves greatly under the application of a transverse magnetic field. Within the IC AF I phase at 6.25 T (second row in Fig. \ref{fig:inelasticoverview}), the higher energy band becomes more diffuse in both {\bf Q} and energy.  The dispersion of the lower energy band is largely unaffected, however the energy gaps have decreased significantly and the spectral weight is weaker. At 7 T (third row in Fig. \ref{fig:inelasticoverview}), in the IC AF II phase, the low energy band near 1 meV has lost much of its low field dispersion and spectral weight at all wavevectors. The higher energy band is now very diffuse and weak. Clearly the incommensurate phases (IC AF I and II) do not support propagating spin wave excitations well. 
As the transverse field increases beyond 7 T, the lower spin wave band gradually acquires more dispersion, and the gap energies evolve differently at different {\bf Q}s, as illustrated in Fig. \ref{fig:gapdependence}(a). The minimum gap of $\sim$ 0.8 meV at {\bf Q}=(2,0,2) shows little transverse field dependence to beyond $\mu_0H_{\rm c3}\sim$ 13 T, however gaps at ${\bf Q}$ = (1,0,0) and (1,0,1) show linear increases with field with the slope of the gap at (1,0,0) vs field being about 50 $\%$ greater than that at (1,0,1).
On passing into the high field Q-PM phase at $\mu_0H_{\rm c3}\sim$ 13 T, the spin waves become sharp and dispersive at all wavevectors in the Brillouin zone, as can be seen at $\mu_0H\parallel b = 15$ T in the bottom row of Fig. \ref{fig:inelasticoverview}. At these high transverse fields, the resolution limited spin wave dispersion is parabolic at all wavevectors for which the dispersion is at a minimum. The very structured spin wave dispersion and intensities are well suited for spin wave theory analysis and extraction of a detailed spin Hamiltonian, although that is beyond the scope of the present work.

The QPT at $\mu_0H_{\rm c3}\sim$ 13 T appears continuous in nature, as incommensurate Bragg intensities, such as that of $(1,\delta, 1)$ go continuously to zero at $H_{\rm c3}$ (see Fig. \ref{fig:intelasticpeaks}(b).  This continuous nature of the QPT is also expressed in the spin dynamics, as seen in Fig. \ref{fig:gapdependence}(c), where constant energy maps in reciprocal space through the low energy branches are shown. The spin excitation spectrum for 1.8meV $<$ E $<$ 2.2 meV is seen to evolve smoothly from $\mu_0H =$ 12 T to $\mu_0H=$ 15 T, with the major difference being greater anisotropy in the $a-c$ plane spin wave velocities above $\mu_0H_{\rm c3}$.  One can also appreciate the distinct eigenvectors associated with the spin excitations in this low energy band by comparing the distribution of inelastic scattering within the (H, 0, L) plane at zero and 15 T transverse field (Fig. \ref{fig:gapdependence}(c)-\textit{(i)}), revealing a shift of the strongest spin wave cones from along [H,0,0] in zero field to along [0,0,L] in 15 T. This is consistent with spin fluctuations along $c$ in zero field, and along $a$ at 15 T, above $\mu_0H_{\rm c3}$. At 15 T, the elastic scattering is consistent with two magnetic subsystems, both polarized normal to $a$; hence their fluctuations can be along the easy $a$ direction, consistent with the {\bf Q}-dependence of the low energy spin excitations in Fig. \ref{fig:gapdependence}(c)-\textit{(iii)}.  

To conclude, we observe three transverse field induced QPTs at $\mu_0H_{\rm c1}\sim6.25$ T, $\mu_0H_{\rm c2}\sim7$ T and $\mu_0H_{\rm c3}\sim 13$ T and 1.6 K in the quasi 2D Ising magnet Co$_3$V$_2$O$_8$.  The relevant transverse field-temperature phase diagram is determined.  A sufficiently high transverse field induces a continuous phase transition to a novel structure in which the spine and cross-tie moments on the kagome staircase lattice decouple, and form AF spin-flopped and transverse field polarized sublattices, respectively. The spin excitation spectra in all relevant transverse field induced phases have been mapped out, and show the evolution of the spin wave eigenvectors with structure.  

We wish to thank L. Balents, M.J.P. Gingras, P. Henelius, O. Petrenko, and L. Savary for helpful discussions. Work at McMaster University was supported by NSERC of Canada. The research at Oak Ridge National Laboratory's Spallation Neutron Source was sponsored by the Scientific User Facilities Division, Office of Basic Energy Sciences, U.S. Department of Energy. The data were reduced using Mantid \cite{mantidproject} and analysed using the HORACE software package \cite{Horace}.

\end{document}

\title{Quantum phase transitions and decoupling of magnetic sublattices in the quasi-two-dimensional Ising magnet Co$_{3}$V$_{2}$O$_{8}$ in a transverse magnetic field (Supplemental Material)}
\author{K. Fritsch}
\affiliation{Department of Physics and Astronomy, McMaster University, Hamilton, Ontario, L8S 4M1, Canada}
\affiliation{Helmholtz Zentrum Berlin f\"{u}r Materialien und Energie GmbH, D-14109 Berlin, Germany}

\author{G. Ehlers}
\affiliation{Quantum Condensed Matter Division, Oak Ridge National Laboratory, Oak Ridge, Tennessee 37831-6475, USA}

\author{K. C. Rule}
\altaffiliation[current address: ]{The Bragg Institute, ANSTO, Kirrawee DC NSW 2234, Australia}
\affiliation{Helmholtz Zentrum Berlin f\"{u}r Materialien und Energie GmbH, D-14109 Berlin, Germany}

\author{K. Habicht}
\affiliation{Helmholtz Zentrum Berlin f\"{u}r Materialien und Energie GmbH, D-14109 Berlin, Germany}

\author{M. Ramazanoglu}
\altaffiliation[current address: ]{Physics Eng. Dept., Science and Letters Faculty, Istanbul Technical University, Maslak, TR-34469, Istanbul, Turkey}
\affiliation{Department of Physics and Astronomy, McMaster University, Hamilton, Ontario, L8S 4M1, Canada}

\author{H. A. Dabkowska}
%\affiliation{Department of Physics and Astronomy, McMaster University, Hamilton, Ontario, L8S 4M1, Canada}
\affiliation{Brockhouse Institute for Materials Research, Hamilton, Ontario, L8S 4M1, Canada}

\author{B. D. Gaulin}
\affiliation{Department of Physics and Astronomy, McMaster University, Hamilton, Ontario, L8S 4M1, Canada}
\affiliation{Brockhouse Institute for Materials Research, Hamilton, Ontario, L8S 4M1, Canada}
\affiliation{Canadian Institute for Advanced Research, 180 Dundas St.\ W.,
Toronto, Ontario, M5G 1Z8, Canada}

%\date{\today}

\maketitle

%\section{Introduction}
This Supplemental Material contains further information related to the data presented in the main manuscript as well as additional data obtained from higher resolution inelastic neutron scattering measurements on the same sample using the cold triple axis spectrometer FLEX at Helmholtz Zentrum Berlin (HZB).

\begin{figure*}
\includegraphics[width=16cm]{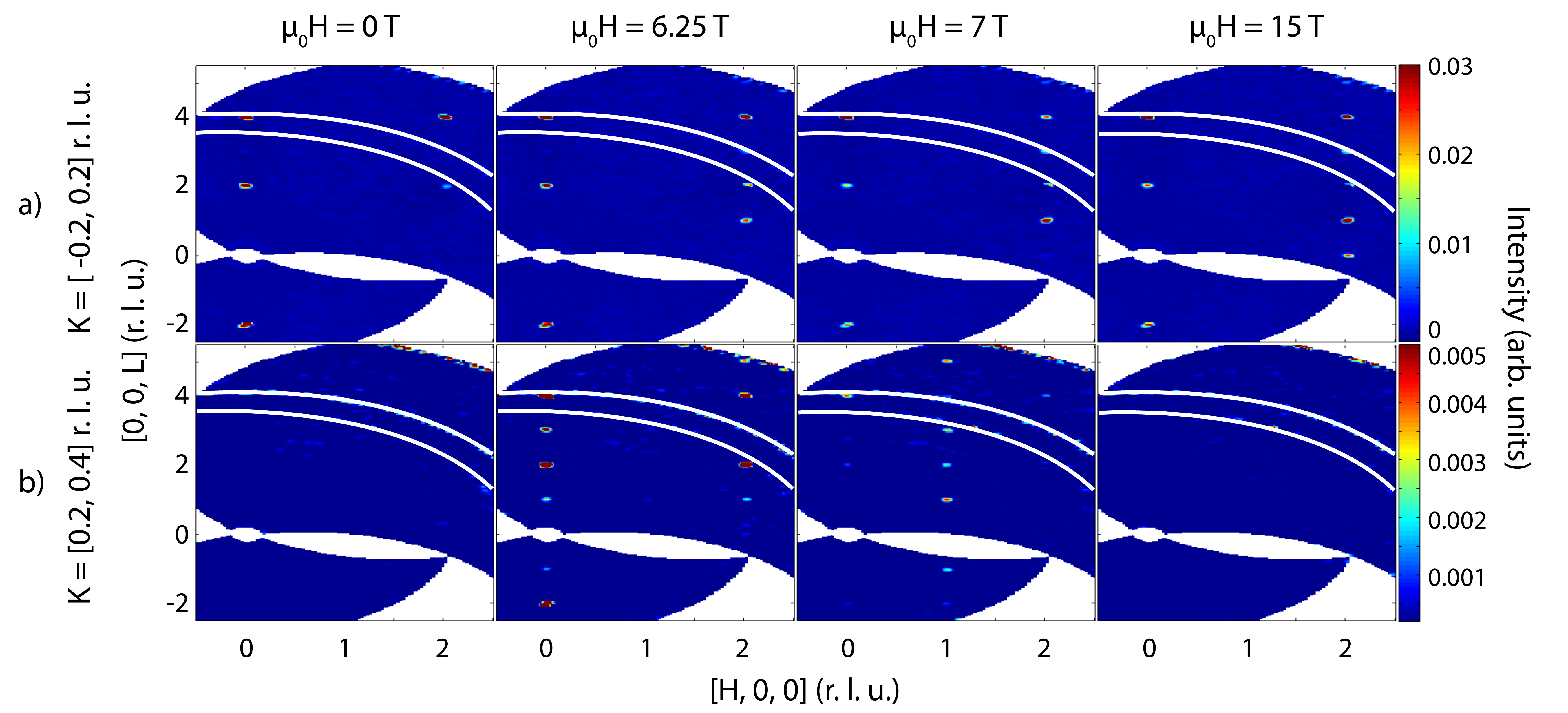}
\caption{(Color online) Elastic scattering for $-0.4<E<0.4$ meV in the horizontal scattering plane $K=[-0.2, 0.2]$ r. l. u. and slightly above the scattering plane in the vertical direction $K = [0.2, 0.4]$ r. l. u.. In an applied transverse magnetic field, the magnetic scattering intensities show a large change at the phase transitions at 6.25 T and 7 T. Note the different intensity scales for the two scattering planes. The scattering out of plane is roughly an order of magnitude weaker than the in-plane scattering. A 30 K data set from within the high temperature paramagnetic phase has been used as nonmagnetic background and subtracted from all data. The data were also corrected for detector efficiency. The white rings denote the powder lines from the aluminum sample mount and environment.}
\label{fig:elasticmaps}
\end{figure*}

Figure \ref{fig:elasticmaps} shows reciprocal space maps of the elastic scattering obtained using CNCS. The vertical angular opening of the magnet cryostat, together with the spectrometer's position-sensitive detectors, allowed us to collect scattering data for positions in reciprocal space of the type $[H,K,L]$ for a limited range of $K = [-0.4,0.4]$ r.l.u. It is therefore possible to separate scattering from within the horizontal plane ($K=[-0.2,0.2]$ r.l.u.), shown in Fig. \ref{fig:elasticmaps}(a), from scattering at positions in the vertical direction ($K=[0.2,0.4]$ r.l.u.), shown in Fig. \ref{fig:elasticmaps}(b). The vertical, $K$, direction is the direction pointing along the applied magnetic field.

The data shown in these scattering maps have been corrected for detector efficiency and have had a high temperature background taken at $T=30$ K, from within the paramagnetic phase, subtracted from it. The high temperature, 30 K, background was taken with the sample in the magnet, and was applied to all data sets, including data in both zero and non-zero magnetic field. The resulting scattering intensities are therefore magnetic in origin.

From these data, the evolution of the magnetic phases in transverse field can be appreciated. In zero field, the in-plane scattering shows strong magnetic intensity at the (0,0,$\pm$2), (0,0,4) and (2,0,4) positions, and lower intensity at (2,0,2), but no scattered intensity is observed in the vertical detector banks ($\delta\sim0.3$), all consistent with the ferromagnetic (F) phase. In a field of 6.25 T, which corresponds to a field just above the transition into the incommensurate antiferromagnetic (IC AFM I) phase, the elastic scattering is characterized by the appearance of the (2,0,1) peak in the horizontal plane, and by the appearance of strong scattering at the (0,$\delta,\pm$1), (0,$\delta,\pm$2), (0,$\delta$,3), (0,$\delta$,4), as well as at the (2,$\delta$,1), (2,$\delta$,2), and (2,$\delta$,4) incommensurate positions with  $\delta\sim0.3$. Upon raising the field to 7 T, and moving into the IC AF II phase, the line of reflections at ($0,\delta,L$) positions disappears and new reflections of type ($1,\delta,L$) with integer $L$ emerge. At the same time, the intensity of the $(2,0,1)$ reflection further increases, while the intensity at $(0,0,2)$ drops significantly compared to that at 6.25 T. In the high-field Q-PM state at 15 T, the incommensurate peaks have disappeared, and the in-plane scattering is dominated by the $(2,0,4)$, $(0,0,2)$ and $(2,0,1)$ reflections, as well as an additional $(2,0,0)$ peak. This scattering is consistent with a magnetic structure in which the cross-tie sites form an antiferromagnetic sublattice polarized along $c$, while the spine site moments are polarized with the applied field along $b$, as shown in Fig. 1(c) of the main manuscript. Based on the relative intensities of the main reflections at (0,0,2), (2,0,4), (2,0,0) and (2,0,1) in the ferromagnetic zero-field phase and the high-field phase, the magnetic moments could be determined as $\sim 3 \mu_{\rm B}$ on the cross-tie sites and $\sim 1\pm0.2 \mu_{\rm B}$ on the spine sites.

Cold triple axis neutron scattering measurements were performed at HZB using the FLEX spectrometer with focussing PG monochromator and analyzer at fixed final wave vectors $k_f$=1.3 \AA$^{-1}$ ($E_f$=3.5 meV) and $k_f$=1.55 \AA$^{-1}$ ($E_f$=4.98 meV) for high resolution measurements at small and large energy transfers, respectively. For these inelastic measurements, the collimation was set to 60'-open-open (using the convention: monochromator to sample-sample to analyser-analyser to detector), and a cooled Be-filter was used on the scattered neutron side of the spectrometer, after the sample, to reduce higher-order wavelength contamination. The sample was mounted in a dilution refrigerator with a base temperature of 50 mK and a 14.5 T vertical field magnet.

\begin{figure}
\includegraphics[width=8cm]{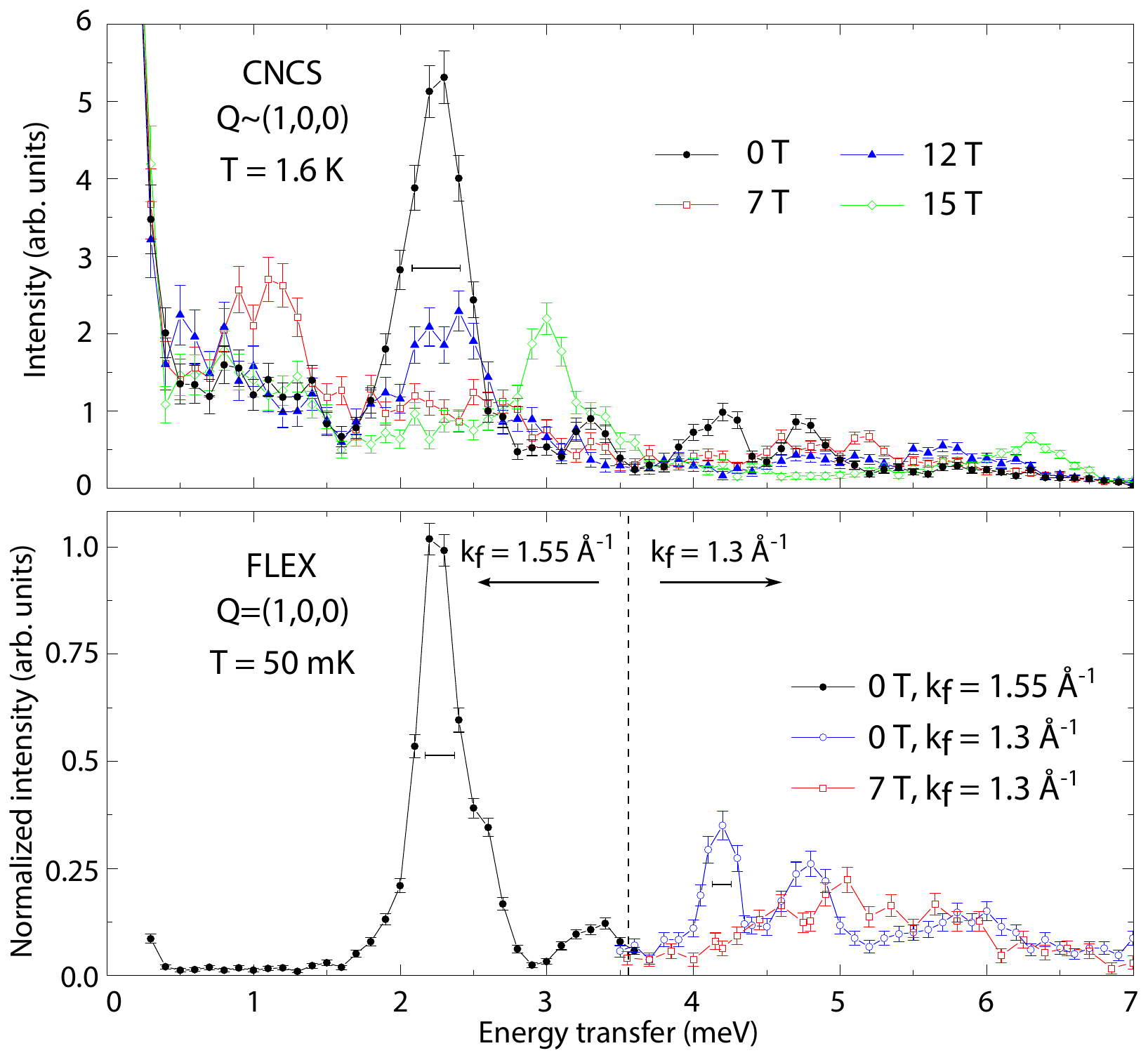}
\caption{(Color online) Comparison of a constant ${\bf Q}$-cut from time-of-flight data from CNCS, SNS, taken at $E_i=12$ meV, and cold triple axis data taken with $E_f=3.5$ ($k_f$=1.3 \AA$^{-1}$) and 4.98 meV ($k_f$=1.55 \AA$^{-1}$) at FLEX, HZB. The respective energy resolutions are shown as black horizontal bars. Data at 1.6 K (CNCS) and 50 mK (FLEX) are both taken at $T\ll T_{\rm C}$ and look qualitatively the same, in zero field and at 7 T. The integration range for the constant ${\bf Q}$-cut at $(1,0,0)$ on CNCS was $\pm0.1$ r. l. u. in $H$ and $L$ and $\pm0.4$ r. l. u. in $K$. CNCS data are corrected for detector efficiency only.}
\label{fig:FLEXCNCS}
\end{figure}

In Fig. \ref{fig:FLEXCNCS}, we show the comparison of a constant-${\bf Q}$ cut at $(1,0,0)$ as measured in two different experimental environments. We compare here data obtained at $T=1.6$ K from the cold neutron chopper spectrometer CNCS (top panel) taken at $E_i=12$ meV and an elastic energy resolution of 0.45 meV with cold triple-axis data measured at a temperature of 50 mK on FLEX (bottom panel) utilizing $k_f$=1.3 \AA$^{-1}$ (elastic resolution of 0.11 meV) and $k_f=1.55$ \AA$^{-1}$ (elastic resolution 0.20 meV) for different ranges in energy transfer. The integration ranges are chosen to approximate the constant-${\bf Q}$ cut at $(1,0,0)$ from the CNCS time-of-flight data. These ranges are $H=1\pm0.1$ r.l.u., $K=\pm0.4$ r.l.u. and $L=\pm0.1$ r.l.u. Apart from a slight broadening due to the instrumental resolution $\delta E$ of CNCS, which is indicated in Fig. \ref{fig:FLEXCNCS} a) by a horizontal bar for the corresponding energy transfer of $\Delta E \sim 2$ meV where $\delta E\sim 0.35$ meV (calculated from \cite{Ehlers2011}), the CNCS data at the higher temperature look qualitatively similar to the data taken on FLEX. In zero field, six modes are observed, four at energy transfers of $3<E<7$ meV and two modes at lower energy transfers of $\sim2$ meV (the lowest energy band shown here at $\sim 2.2$ meV actually contains two branches as shown below). At the phase transition at 7 T, the low energy band is suppressed in energy, while the higher energy modes become weak and diffuse with a width of $\gtrsim 1.5$ meV, typical of incommensurate or disordered systems with very short lifetimes on the order of picoseconds. At higher fields, a reorganization of the spin excitations takes place and sharp resolution-limited spin waves develop. Moreover, the similarity between the 50 mK and 1.6 K data implies that thermal fluctuations play little role in the evolution of the phases and their spin fluctuations.  At $T=1.6$ K we are truly probing the transverse field-induced quantum fluctuations in this system. This is not so surprising given that 1.6 K is deep within the $H=0$ ferromagnetic ground state and is separated from intermediate incommensurate phases by a strongly first order phase transition at $T_{\rm C}=6.2$ K.  

\begin{figure}
\includegraphics[width=8cm]{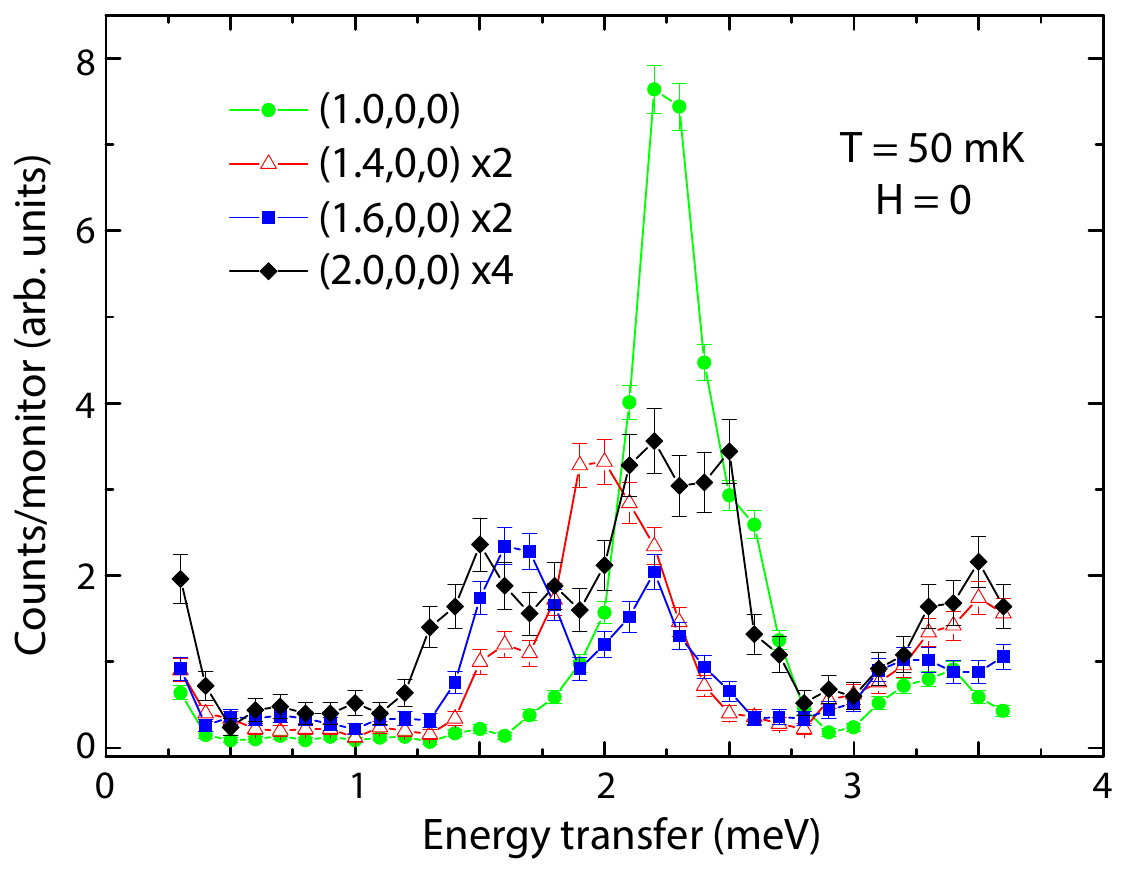}
\caption{(Color online) Constant-${\bf Q}$ cuts at 50 mK and $H=0$ obtained at FLEX with $E_f=3.5$ meV, showing the evolution from {\bf Q}=(1.0,0,0) to {\bf Q}=(2.0,0,0). Note that the intensities of the {\bf Q}=(1.4,0,0) and {\bf Q}=(1.6,0,0) scans are multiplied by 2, that of the {\bf Q}=(2.0,0,0) scan multiplied by a factor of 4.}
\label{fig:suppl3}
\end{figure}

Figure \ref{fig:suppl3} shows the evolution of the zero-field low energy modes for constant-{\bf Q} scans ranging from {\bf Q}=(1.0,0,0) to {\bf Q}=(2.0,0,0), that is moving from the ferromagnetic Brillouin zone boundary to the Brillouin zone center. One can clearly see that the apparent single mode around 2.25 meV at {\bf Q}=(1.0,0,0) splits into two modes with energies 1.5 and 2.25 meV at {\bf Q}=(2.0,0,0). These modes carry much less spectral weight (roughly a factor of 8 less intensity) than those at {\bf Q}=(1.0,0,0), which is why they were not clearly observed in the CNCS data.     

We wish to thank K. Kiefer and S. Gerischer for providing the low temperature sample environment for the FLEX measurements at HZB.